\documentclass[showpacs,showkeys,prb,twocolumn]{revtex4}
\usepackage{amsfonts}
\usepackage{graphicx}
\usepackage{amsmath}
\usepackage{amssymb}%
\begin{document}

\title[Divergent beams]{Divergent beams of nonlocally entangled electrons emitted from hybrid
normal-superconducting structures}
\author{Elsa Prada}
\affiliation{Departamento de F\'{\i}sica Te\'{o}rica de la Materia Condensada, Universidad
Aut\'{o}noma de Madrid, E-28049 Madrid, Spain, and Institut f\"{u}r
Theoretische Festk\"{o}rperphysik, Universit\"{a}t Karlsruhe, 76128 Karlsruhe, Germany}
\email{elsa@tfp.uni-karlsruhe.de}
\author{Fernando Sols}
\affiliation{Departamento de F\'{\i}sica de Materiales, Facultad de
Ciencias F\'{\i}sicas, Universidad Complutense de Madrid, E-28040
Madrid, Spain} \email{f.sols@fis.ucm.es} \keywords{entanglement,
resonant tunneling, hybrid structures, heterostructures, electron
transport\\ Published in: New J. Phys. \textbf{7}, 231 (2005)}

\begin{abstract}
We propose the use of normal and Andreev resonances in normal-superconducting
structures to generate divergent beams of nonlocally entangled electrons.
Resonant levels are tuned to selectively transmit electrons with specific
values of the perpendicular energy, thus fixing the magnitude of the exit
angle. When the normal metal is a ballistic two-dimensional electron gas, the
proposed scheme guarantees arbitrarily large spatial separation of the
entangled electron beams emitted from a finite interface. We perform a
quantitative study of the linear and nonlinear transport properties of some
suitable structures, taking into account the large mismatch in effective
masses and Fermi wavelengths. Numerical estimates confirm the feasibility of
the proposed beam separation method.

\end{abstract}
\volumeyear{year}
\volumenumber{number}
\issuenumber{number}
\eid{identifier}
\startpage{1}
\endpage{ }
\maketitle

\section{Introduction}

\label{Intro} The goal of using entangled electron pairs for the processing of
quantum information poses a technological challenge that requires novel ideas
on electron quantum transport. It has been proposed that a conventional
superconductor is a natural source of entangled electrons which may be emitted
into a normal metal through a properly designed interface
\cite{rech01,leso01,rech02,chtc02,fein03,rech03,samu03,prad04,samu04,meli04,saur04}%
. At low temperatures and voltages, the electric current through a
normal-superconducting (NS) interface is made exclusively of electron Cooper
pairs whose internal singlet correlation may survive for some time in the
context of the normal metal. The emission of two correlated electrons from a
superconductor into a normal metal is often described as the Andreev
reflection \cite{andr64} of an incident hole which is converted into an
outgoing electron. The equivalence between the two pictures has been
rigorously proved in Refs. \onlinecite{samu03,prad04,samu05}. There the
relation was established between the various quasiparticle scattering channels
as these are referred to different choices of normal metal chemical potential,
i.e. to different definitions of the vacuum. When the reference chemical
potential employed to label quasiparticle states in the normal metal is
identical to the superconductor chemical potential ($\mu_{N}=\mu_{S}$), the
number of Bogoliubov quasiparticles is conserved and the Andreev picture
holds. If, on the contrary, $\mu_{N}$ is chosen to be smaller than $\mu_{S}$,
quasiparticle number conservation is not guaranteed and spontaneous emission
of two electrons through the SN interface becomes possible \cite{prad04}.
Transport calculations across an SN interface at low temperature and voltage
which invoke an explicit two-electron picture have been presented in Refs. \onlinecite{rech01,prad04,hekk93}.

The need for spatial separation of the entangled beams has motivated the
search for schemes that constrain (or at least allow) the two pair electrons
to be emitted from different locations at the NS interface \cite{rech01}. In
the conventional picture where quasiparticle scattering is unitary, that
process is viewed as the absorption of a hole and its subsequent reemission as
an electron from a distant point. Such a crossed (or nonlocal) Andreev
reflection has been observed experimentally \cite{beck04,russ05,aron05}.

The requirement of physical separation is a severe limitation in
practice, since pairing correlations decay with distance. As a
consequence, the current intensity of nonlocally entangled electrons
decreases with the distance $r$ between the two emitting points.
There is an exponential decay on the scale of the superconductor
coherence length which reflects the short-range character of the
superconductor pairing correlations \cite{rech01,prad04}. A more
important limitation in practice comes from the prefactor, which,
besides oscillating on the scale of the superconductor Fermi
wavelength, decreases algebraically with distance. In the tunneling
limit, and for a ballistic 3D superconductor, the decay law is
$r^{-2}$, if the tunneling matrix elements are assumed to be
momentum independent \cite{rech01}, or $r^{-4}$, if proper account
is taken of the low-momentum hopping dependence
\cite{prad04,houz05}. Within the context of momentum-independent
tunneling models, the power law changes if the superconductor is low
($d$) dimensional \cite{rech02,bouc03}, or diffusive
\cite{fein03,bign04}, yielding $r^{-d+1}$ and $r^{-1}$,
respectively. It remains to be investigated how that behavior
changes when more realistic tunnel matrix elements are employed
\cite{prad04,houz05} and when geometries other than planar or
straight boundaries are considered.

In this article we propose an experimental setup that would guarantee long
term separation of correlated electron pairs without the shortcomings caused
by the need to emit the pair electrons from distant points. The idea is
\textit{to transmit both electrons through the same spatial region but
inducing them to leave in different directions}. In a ballistic normal metal
such as a two-dimensional electron gas (2DEG), that divergent propagation
guarantees the long term separation of the entangled electrons at distances
from the source much greater than the size of the source.

To force the pair electrons to leave in different directions, we
propose to exploit the formation of resonances in a properly
designed normal-superconductor interface. These could be
one-electron (normal) resonances, such as those found in
double-barrier structures \cite{chan74} (SININ structure), or
two-electron (Andreev) resonances such as the de Gennes --
Saint-James resonances appearing in structures with one barrier
located on the normal metal side at some distance from the
transmissive SN interface (SNIN structure)
\cite{dege63,ried93,giaz01,giaz03}. Those quasi-bound states have it
in common that, in a perfect interface, they select the
perpendicular energy of the exiting electrons while ensuring the
conservation of the momentum parallel to the interface. At low
voltages and temperatures, this also determines the parallel energy,
given that the total energy of the current contributing electrons is
constrained to lie close to the normal Fermi level. Altogether, this
mechanism fixes the magnitude of the exit angle, since the parallel
momenta of the pair electrons are opposite to each other and both
remain unchanged during transmission through the perfect interface.
Thus the electron velocities form a V-shaped beam centered around
the perpendicular axis.

The type of structures which are needed seems to be within the reach of
current experimental expertise. In the last fifteen years, several groups have
built a variety of hybrid superconductor-semiconductor (SSm) structures
\cite{giaz01,giaz03,akaz91,kast91,gao93,nguy94,tabo96,defr98,russ05}. More
recently, some experimental groups \cite{toyo99,bato04,choi05} have
investigated transport through SSm structures where Sm is a 2DEG on a plane
essentially perpendicular to the superconductor boundary. In such setups, the
SN interface lies at the one-dimensional (1D) border of the two-dimensional
(2D) ballistic metal. If two parallel straight-line barriers were drawn in
that structure, one along the SN interface and another one at some distance
within N, then the experimental scenario considered in this article would be
reproduced. \ A three-dimensional (3D) version of the same structure, in which
Sm would be 3D and the interface would be 2D, of the type reported in Ref.
\onlinecite{giaz03}, would also produce divergent electron beams. These,
however, would be emitted into a 3D semiconductor, where it may be more
difficult to pattern suitable detectors.

Once the two electrons propagate in the ballistic 2DEG, their motion can be
controlled by means of existing techniques. For instance, they can be made to
pass through properly located narrow apertures, such as those used in electron
focusing experiments \cite{vanh88}. For quantum information processing, their
spin component in an arbitrary direction could eventually be measured by using
the Rashba effect \cite{rash60,datt90} to rotate the spin before electrons
enter the spin filter \cite{schl03}. Then one could attempt to measure Bell
inequalities \cite{leso01,chtc02,samu03,been03,faor04,saur05,prad05,tadd05}.
Alternatively, one may measure electric current cross-correlations
\cite{samu04,torr99,burk00,samu02,bign04} to indirectly detect the presence of
singlet spin correlations.

In Section II, we describe the model we have adopted for our
calculations. Two important features are the offset between the
conduction band minima and the difference in the effective masses of
S and Sm. Both effects have been analyzed by Mortensen \textit{et
al.} \cite{mort99} in the context of SIN structures, with N a 3D
semiconductor. In Section III, we focus on the linear regime and
calculate the zero bias conductance using the multimode formula
derived by Beenakker \cite{been92}. There we investigate the angular
distribution of the outgoing electron current and observe how it is
indeed peaked around two symmetric directions. Section IV is devoted
to the nonlinear regime \cite{leso97}, where the voltage bias may be
comparable to the superconductor gap. We find divergent beams again,
this time with new features caused by the difference between the
electron and hole wavelengths. By plotting the differential
conductance, we relate our work to the previous literature on SN
transport and note the presence of a reflectionless tunneling zero
bias peak \cite{giaz03,kast91,mels94}, as well as the existence of
de Gennes -- Saint-James resonances. In Section V, we discuss how
the need to have a broad perfect interface, as required for parallel
momentum conservation, can be reconciled with the interface finite
size which is needed for the eventual spatial separation of the
emerging beams. We conclude in Section VI.

\section{The model}

\label{model}

\begin{figure}
[ptb]
\includegraphics[width=7cm]{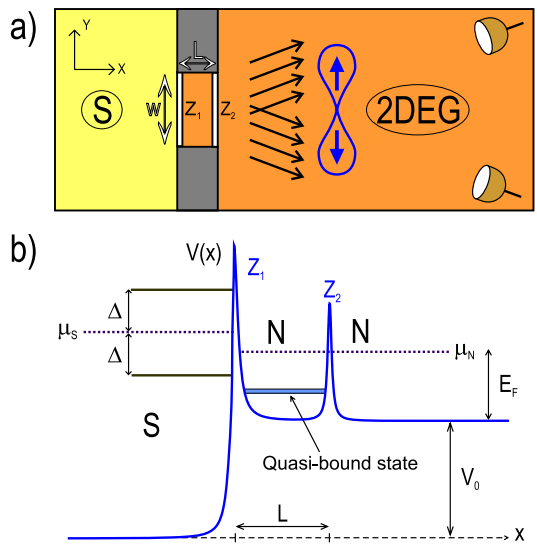}
\caption{(Color online) (a) Schematic representation of the SN
structure considered in this article. S is a conventional
superconductor; N is a two-dimensional electron gas. Energy
selection in the $x$ direction by the resonance structure limits
current flow to two divergent beams of entangled electrons. (b)
One-electron potential profile in the $x$ direction. Quasi-bound
electron states form due
to multiple reflection by two barriers of strength $Z_{1}$ and $Z_{2}$.}%
\label{system}%
\end{figure}

We wish to investigate the role of resonances in the angular distribution of
the normal current in suitably designed SSm interfaces. A prototypical
structure is shown in Fig. 1a, where the 2DEG forms an angle with the planar
boundary of a superconductor, similar to the setup built in Ref. \onlinecite{choi05}.

In the present analytical and numerical work we consider a semi-infinite
ballistic 2DEG (hereafter also referred to as N) lying in the half-plane
$x>0$. We assume a perfect interface, so that the one-electron potential is
independent of $y$. Specifically, $V(x)$ is taken of the form%
\begin{equation}
V(x)=-V_{0}\Theta(-x)+H_{1}\delta(x)+H_{2}\delta(x-L)\ . \label{potential}%
\end{equation}
Here, $V_{0}$ accounts for the large difference between the widths of the S
and N conduction bands. If $E_{F}=\hbar^{2}k_{F}^{2}/2m$ and $E_{F}^{\prime
}=\hbar^{2}k_{F}^{\prime2}/2m^{\prime}$ are N and S Fermi energies,
respectively, one typically has $E_{F}^{\prime}\sim V_{0}\gg E_{F}\gg\Delta$,
where $\Delta$ is the zero-temperature superconducting gap. We assume that the
bulk parameters change abruptly at $x=0$. The structure contains two delta
barriers, located at the SN interface and at a distance $L$ from it within the
N side. Their reflecting power is measured by the dimensionless parameters
$Z_{1}$ and $Z_{2}$, defined as $Z_{1}=H_{1}/\hbar\left(  v_{F}v_{F}^{\prime
}\right)  ^{1/2}$ and\ $Z_{2}=H_{2}/\hbar v_{F}$. The effective mass $m$, the
Fermi wavevector $k_{F}$, and the Fermi velocity $v_{F}$ are those of the
normal 2DEG, while $m^{\prime}$, $k_{F}^{\prime}$, and $v_{F}^{\prime}$
correspond to a conventional superconductor.

It was shown in Ref. \onlinecite{prad04} that the picture of two-electron
emission and hole Andreev reflection are equivalent. For computational
purposes, we employ here the standard Andreev picture whereby all
quasiparticles have positive energy ($\varepsilon>0$), with the quasiparticle
energy origin given by $\mu_{S}$. However, in our discussion we will
occasionally switch between the two images. An important feature is that the
absence of a hole at $\varepsilon>0$ in the Andreev scenario corresponds to
the presence of an electron at $-\varepsilon<0$ in the two-electron picture
\cite{prad04}.

In a transport context, the superconductor and normal metal chemical
potentials differ by $\mu_{S}-\mu_{N}=eV$, where $V$ is the applied bias
voltage. In the Andreev picture, one artificially takes $\mu_{S}$ as the
reference chemical potential for labeling quasiparticles and the imbalance
$eV$ is accounted for by introducing an extra population of incoming holes
with energies between 0 and $eV$ \cite{blon82,lamb91}.

An apparent shortcoming of the Andreev picture is that it does not
show explicitly that the emitted electron pairs are internally
entangled. In this respect, we may note the following remarks:\ (i)
the two-electron hopping matrix element vanishes when the spin state
in the N side is a triplet \cite{rech01};\ (ii) an analytical study
of transport through a broad SN interface based on a two-electron
tunneling picture \cite{prad04} (with the final state explicitly
entangled) gives results identical to those obtained within an
Andreev description \cite{kupk97};\ (iii) entanglement in the
outgoing electron pairs has been explicitly proven in the general
tunneling case \cite{samu05}; and (iv) transport across the SN
structure is spin independent and thus must preserve the internal
spin correlations of the emitted electron pair \cite{oh05}.
Moreover, using full counting statistics Samuelsson \cite{samu03a}
has shown that current through an SN double-barrier structure is
carried by correlated electron pairs.

To compute the current, we must sum over momenta parallel to the
interface, which on the N side take values $-k_{F}<k_{y}<k_{F}$. For
the purposes of solving the one-electron scattering problem, we
assume that the superconductor is also two-dimensional. Due to the
mismatch in effective masses, the perpendicular energy is not
conserved (refraction). The conserved quantum numbers are the
parallel momentum ($k_{y}=k_{y}^{\prime}$) and the total energy
($E_{x}+E_{y}=E_{x}^{\prime}+E_{y}^{\prime}-V_0=E$, with $E_{x}\neq
E_{x}^{\prime}$).

For a given $k_{y}$, the energy available for perpendicular motion is
$E_{x}=E-\hbar^{2}k_{y}^{2}/2m$, where $E$ is the electron total energy. As a
consequence, for each $k_{y}$ the picture depicted in Fig. 1b holds provided
that the $\mu_{N}$ is replaced by an effective value \cite{mort99}
\begin{equation}
\mu_{N}(k_{y})=\mu_{N}-\hbar^{2}k_{y}^{2}/2m\ ,\ \ \label{mu-eff}%
\end{equation}
which is matched to $\mu_{S}(k_{y})=\mu_{S}-\hbar^{2}k_{y}^{2}/2m^{\prime}$,
with $\mu_{S}(k_{y})-\mu_{N}(k_{y})$ generally not equal to $eV$.

Beenakker \cite{been92} has computed the SN zero bias conductance for an
interface with many transverse modes. Mortensen \textit{et al.} \cite{mort99}
have adapted the work of Ref. \onlinecite{blon82} to account for the full 3D
motion through a perfect, 2D SSm interface, where the effective masses and the
Fermi wavelengths of N and S may differ widely. Lesovik \textit{et al.}
\cite{leso97} have generalized the work of Refs. \onlinecite{been92,blon82} to
the nonlinear case where $eV$ may be comparable to $\Delta$. They have applied
their results to structures displaying quasiparticle resonances. Here we
combine the work of these previous three references. Specifically, we
investigate the transport properties of an SN interface for \textit{arbitrary
bias} $V$ between 0 and $\Delta$. We consider structures displaying
\textit{resonances} due to multiple quasiparticle reflection, and allow for a
\textit{large disparity between the }S \textit{and} N \textit{bulk
properties}. Most importantly, we calculate the \textit{angular distribution}
of the pair electron current emitted into the semiconductor. Another novel
feature is that the semiconductor we consider is a 2DEG whose plane forms an
angle with the superconductor planar boundary, so that the SN interface is
formed by a straight line.

\section{Zero bias conductance}

\label{ZBC}

The zero bias conductance is defined as%
\begin{equation}
G(0)\equiv\lim_{V\rightarrow0}dI/dV\ , \label{zbc}%
\end{equation}
where $I$ is the total current at voltage bias $V$. For an SN interface
\cite{been92},%
\begin{equation}
G(0)\equiv\frac{4e^{2}\ }{h}\sum_{\nu=1}^{N}\frac{T_{\nu}^{2}}{\left(
2-T_{\nu}\right)  ^{2}}\ , \label{been}%
\end{equation}
where \{$T_{\nu}$\} are the eigenvalues of the one-electron transmission
matrix through the normal state structure at total energy $E=\mu_{N}\simeq
\mu_{S}\equiv\mu$, and $N$ is the number of transverse channels available for
propagation in the normal electrode at energy $\mu$. For a perfect interface,
the index $\nu$ runs over the possible values of $k_{y}$. Thus, when needed,
we make the replacement $\sum_{\nu}\rightarrow(w/2\pi)\int dk_{y}$, where
$w\rightarrow\infty$ is the interface length. The minimum energy required for
propagation in mode $\nu$, referred to the bottom of the conduction band, is
$\epsilon_{\nu}\equiv\hbar^{2}k_{y}^{2}/2m$.

In the linear regime, the total energy is restricted to be at $\mu$.
Therefore, the running value of $k_{y}$ determines the exit angle%
\begin{equation}
\theta\equiv\arctan\left(  k_{y}/k_{x}\right)  \ , \label{theta}%
\end{equation}
since $k_{x}$ and $k_{y}$ must satisfy
\begin{equation}
k_{x}^{2}+k_{y}^{2}=k_{F}^{2}\ . \label{constraint}%
\end{equation}
Therefore, Eq. (\ref{been}) may be written as%
\begin{equation}
G(0)=\int_{-\pi/2}^{\pi/2}d\theta~G(0,\theta)\ , \label{g-total}%
\end{equation}
with $G(0,\theta)$ properly defined as the \textit{angular distribution} of
the zero bias conductance.

\begin{figure}
[ptbptb]
\includegraphics[width=8cm]{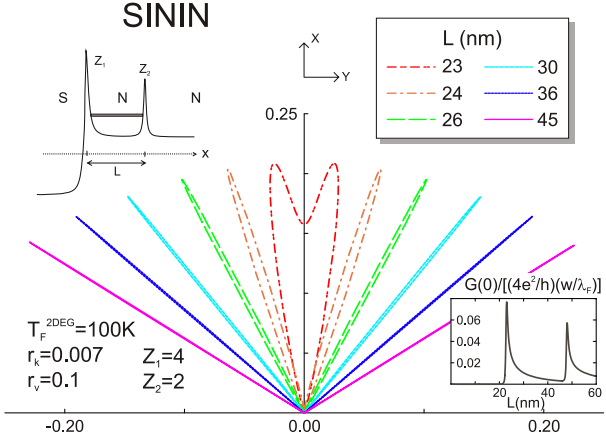}
\caption{(Color online) Normalized angular distribution of the zero
bias conductance for several values of the interbarrier distance
$L$. The barrier strengths are $(Z_{1},Z_{2})=(4,2)$. Upper inset:
schematic representation of the perpendicular potential profile.
Lower inset: total conductance, normalized to half the maximum
possible conductance, as a function of $L$. See main text for
the meaning of other parameters.}%
\label{been1}%
\end{figure}

In Fig. 2, we show $G(0,\theta)$ for several values of the
interbarrier distance $L$, on a structure with potential barriers of
strength $Z_{1}=4$ and $Z_{2}=2$ located at $x=0$ and $L$,
respectively. It is divided by $4e^{2}w/h\lambda_{F}$ ($\lambda_{F}$
being the N Fermi wavelength), which is half the maximum possible
value of $G(0)$ (obtained when $T_{\nu}=1$ for all $\nu$).

The semiconductor conduction band width is taken $E_{F}=k_{B}\times100$ K. The
ratios between the Fermi wavevectors and Fermi velocities in N and S are,
respectively, $r_{k}\equiv k_{F}/k_{F}^{\prime}=0.007$ and $r_{v}\equiv
v_{F}/v_{F}^{\prime}=0.1$ (GaAs values). The presence of quasi-bound states
located between the two barriers yields a structure of resonance peaks in the
one-electron transmission probability $T_{\nu}$ as a function of $\nu$. We
also note that the small value of $r_{k}$ will cause important internal
reflection of the electrons within the superconductor. As a result, only S
electrons very close to normal incidence will have a chance to be transmitted
into N. Once in N, they may leave with much larger angles. Specifically, if
$\theta^{\prime}$ is the angle on the S side, one has $\sin\theta^{\prime
}=r_{k}\sin\theta$ (Snell law). For the parameters considered in this article,
only electrons arriving from S within $\Delta\theta^{\prime}/2=\arcsin
(r_{k})\simeq0.4$ degrees of normal incidence are transmitted through the
normal-state structure.

As $L$ increases, the position of the resonant levels is lowered. In Fig. 2,
the values of $L$ are chosen such that only the lowest resonant level plays a
role. This allows us to investigate the effect of a resonant level at
perpendicular energy (on the N side) $E_{x}=E_{R}\alt\mu$, which appears as a
peak in $T_{\nu}$ as a function of $\nu$. This occurs for $\nu=\nu_{R}$
satisfying%
\begin{equation}
\mu-\epsilon_{\nu_{R}}=E_{R}~. \label{resonancia-normal}%
\end{equation}

For the shortest interbarrier distance displayed ($L=23$ nm), the structure of
$G(0,\theta)$ begins to reveal the presence of a resonance just below $E_{F}$.
The trend towards a bifurcation of the conductance angular distribution
becomes clearer for larger values of $L$. As discussed before, the presence of
a sharp resonance only permits the transmission of electrons with
perpendicular energy $E_{x}$ close to $E_{R}$. This fixes the value of $k_{x}$
at $k_{x}=k_{R}$ and, with it, the magnitude of the exit angle
\begin{equation}
\theta_{R}=\arctan\left(  \sqrt{(k_{F}/k_{R})^{2}-1}\right)  \ .
\label{theta-R}%
\end{equation}
For a given linewidth $\Gamma$ of the one-electron resonance, the
corresponding spread of the angular distribution is%
\begin{equation}
\Delta\theta\simeq\frac{\Gamma}{E_{F}\sin(2\theta_{R})}~, \label{delta-theta}%
\end{equation}
Thus, the angular width has a minimum at $\theta_{R}=\pi/4$, as in fact
revealed by the narrower spikes in Fig. 2.

The lower-right inset of Fig. 2 shows the total conductance [see Eq.
(\ref{g-total})] as a function of the interbarrier distance. It is normalized
to half its maximum possible value. For small $L$, the lowest resonance lies
at $E_{R}>\mu$, which blocks current flow. As $L$ is increased, $E_{R}$
decreases and the lowest resonance becomes available for transport ($E_{R}%
<\mu$). Then $G(0)$ shows a rapid increase followed by a decaying tail. The
effect is so marked that, if we attempt to plot $G(0,\theta)$ for e.g. $L=22$
nm (just below the smallest shown value), the resulting curve is invisible on
the scale of Fig. 2. As $L$ increases further, a second resonance becomes
available for transmission and the wide spikes due to the the first resonance
coexist with the new, more centered lobes which in turn tend to bifurcate as
$L$ increases even more (not shown).

The decay of $G(0)$ for $L>L_{R}$ (where $L_{R}$ is the interbarrier distance
at which $E_{R}=\mu$) goes like $L^{-1/2}$ because it reflects the 1D nature
of the transverse density of states. This can be proved by noting that Eq.
(\ref{been}) can be written as%
\begin{equation}
G(0)=\frac{4e^{2}\ }{h}\sum_{\nu}A_{\nu}~~, \label{sum-transverse}%
\end{equation}
where $A_{\nu}=T_{\nu}^{2}/(2-T_{\nu})^{2}$ is the probability for Andreev
reflection in mode $\nu$ at total energy $\mu$, which corresponds to
quasiparticle energy $\varepsilon=0$. Because of the normal resonance, both
$T_{\nu}$ and $A_{\nu}$ are strongly peaked around the value of $\nu_{R}$
satisfying (\ref{resonancia-normal}). Thus we may approximate $A_{\nu}\simeq
a\delta(\mu-\epsilon_{\nu}-E_{R})$, where $a$ is an appropriate weight. Then
$G(0)$ becomes%
\begin{equation}
G(0)\simeq\frac{4e^{2}}{h}aD(\mu-E_{R})\ , \label{transverse}%
\end{equation}
where $D(\epsilon)\equiv\sum_{\nu}\delta(\epsilon-\epsilon_{\nu})$ is the
transverse density of states. On this energy scale, $E_{R}$ is a smooth
function of $L$, so that it can be approximated as $E_{R}\simeq\mu-b(L-L_{R}%
)$, with $b>0$. Then Eq. (\ref{transverse}) yields $G(0)\propto D(b(L-L_{R}%
))\sim(L-L_{R})^{-1/2}$, as observed in the inset of Fig. 2. Such a
manifestation of the transverse density of states in the total transport
properties is characteristic of structures which select the energy in the
propagation perpendicular to the plane of the heterostructure \cite{wagn99}.
The foregoing argument allows us to predict that, for a 3D structure, the
total conductance will display steps as a function of $L$, since then
$D(\epsilon)$ will be constant (not shown).

\begin{figure}
[ptbptbptb]
\includegraphics[width=8cm]{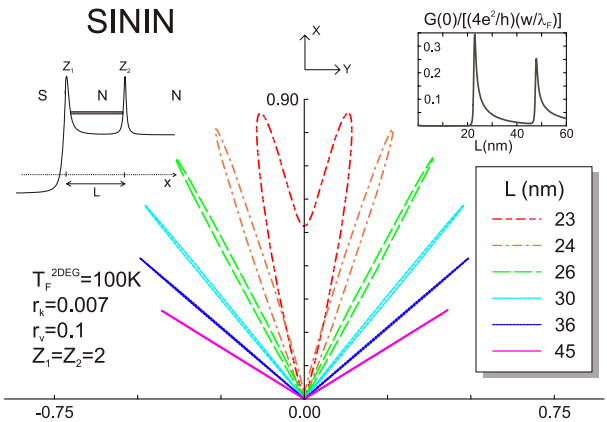}
\caption{(Color online) Same as Fig. \ref{been1}, for $(Z_{1},Z_{2})=(2,2)$. }%
\label{been2}%
\end{figure}

\begin{figure}
[ptbptbptbptb]
\includegraphics[width=8cm]{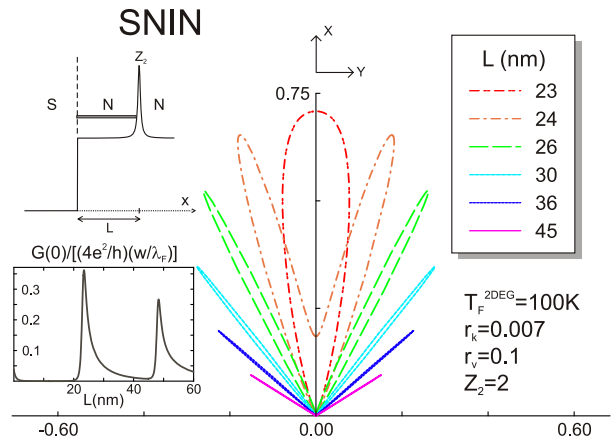}
\caption{(Color online) Same as Fig. \ref{been1}, for $(Z_{1},Z_{2})=(0,2)$. }%
\label{been3}%
\end{figure}

Figs. 3 and 4 show $G(0,\theta)$ for setups identical to that of Fig. 2,
except for $Z_{1}$ taking values 2 and 0, respectively, $Z_{2}$ remaining
fixed at 2. The building of SSm interfaces with small $Z_{1}$ seems feasible
with the doping techniques implemented in Refs. \onlinecite{kast91,tabo96,giaz03}.

As in Fig. 2, the electron flow is channeled through well-defined
resonances in the $x$ direction, again giving rise to divergent
beams in the N electrode. At first sight it may seem surprising that
for $Z_{1}=0$ one still finds peaks in the angular distribution,
since they reveal a structure in the transmission $T_{\nu}$ that is
not expected from a single barrier of strength $Z_{2}$. However,
when $Z_{1}=0$, there is still some normal reflection at $x=0$ due
to the large mismatches $E_{F}^{\prime}\gg E_{F}$ and $m^{\prime}\gg
m$. In fact, on quite general grounds, one has $T_{\nu}\rightarrow0$
as $\epsilon_{\nu}\rightarrow E_{F}$ (equivalent to
$k_{x}\rightarrow0$), even if $Z_{1}=0$. This trend is revealed by
the decreasing length of the spikes for increasing $\theta$
(decreasing $k_{x}$).

In Fig. 3, $\mu$ stays slightly above $E_{R}$ for $L=$ 23 nm. The details of
reflection at the interface cause some shift in the detailed position of the
resonances. For $Z_{1}=0$ (Fig. 4), the resonant level $E_{R}$ at that
particular interbarrier distance is exactly at $\mu$, as revealed by the
absence of splitting in $G(0,\theta)$. If, by decreasing $L$, $E_{R}$ were
taken considerably above $\mu$, then the forward lobe of Fig. 4 would be
sharply reduced. This general property was already noted in the discussion of
Fig. 2 and its inset.\bigskip

\section{Nonlinear transport:\ spectral conductance}

\label{NLDC}

We have seen that, in the zero bias limit, the peaks in the angular
distribution directly reflect the structure of (normal) resonances
in $T_{\nu }$ as a function of $\nu$, since this determines $G(0)$
through Eq. (\ref{been}). As $V$ becomes nonzero and comparable to
$\Delta$, new resonances appear which are a direct manifestation of
Andreev reflection occurring at nonzero quasiparticle energies. Such
Andreev resonances have been discussed, for instance, in Refs.
\onlinecite{ried93,leso97,giaz03}. Below we present a brief
description that suits our present needs and which complements the
discussion given by Lesovik \textit{et al.} \cite{leso97}.

We restrict our study to the case $0<\left\vert eV\right\vert <\Delta$. As in
Ref. \onlinecite{leso97}, we focus for simplicity on the spectral conductance
$G(\varepsilon,V)$, i.e. we neglect the contribution to the total differential
conductance coming from the derivative with respect to $V$ of $G(\varepsilon
,V)$ itself. From Ref. \onlinecite{leso97}, we note that, for $0<\left\vert
\varepsilon\right\vert <\Delta$,%
\begin{widetext}
\begin{align}
G(\varepsilon,V)  &  =\frac{4e^{2}}{h}\sum_{\nu}g_{\nu}(\varepsilon
,V)~,\label{g-gene}\\
g_{\nu}(\varepsilon,V)  &  \equiv\frac{T_{\nu}(\varepsilon)T_{\nu
}(-\varepsilon)}{1+R_{\nu}(\varepsilon)R_{\nu}(-\varepsilon)-2\left[  R_{\nu
}(\varepsilon)R_{\nu}(-\varepsilon)\right]  ^{1/2}\cos\left[  \varphi_{\nu
}(\varepsilon)-\varphi_{\nu}(-\varepsilon)-2\vartheta(\varepsilon)\right]
}~.~ \label{reduced-g}%
\end{align}
\end{widetext}
Here, $g_{\nu}(\varepsilon,V)$ is the Andreev
reflection probability for a quasiparticle of energy $\varepsilon$
incoming in mode $\nu$, with $\left\vert \varepsilon\right\vert
<|eV|$. It is determined by $T_{\nu}(\varepsilon)$, which is defined
as the transmission probability for an electron incident from the N
side on the normal structure (i.e. with \ $\Delta=0$) in transverse
mode $\nu$ with total energy $\mu_{S}+\varepsilon$,
$R_{\nu}(\varepsilon)=1-T_{\nu }(\varepsilon)$,
$\vartheta(\varepsilon)\equiv\arccos(\varepsilon/\Delta)$, and
$\varphi_{\nu}(\varepsilon)$ is the phase of the reflection
amplitude for an electron impinging from the S side on the normal
structure. The latter depends on $\varepsilon$ through the phases
acquired upon reflection on each barrier (usually negligible) and,
more importantly, through the optical path
between the two barriers $k_{\nu}(\varepsilon)L$, where%
\begin{align}
k_{\nu}(\pm\varepsilon)  &  =\left[  2m(E_{F}+eV-\epsilon_{\nu}\pm
\varepsilon)\right]  ^{1/2}/\hbar\nonumber\\
&  \simeq mv_{F\nu}/\hbar+(eV\pm\varepsilon)/\hbar v_{F\nu}~.
\label{k-nu-eps2}%
\end{align}
Here, $v_{F\nu}=\left[  2\left(  E_{F}-\epsilon_{\nu}\right)  /m\right]
^{1/2}$ is the perpendicular velocity for a Fermi electron in mode $\nu$ on
the N side (note that $E_{F}+eV$ is the energy difference between the S
chemical potential and the bottom of the N conduction band). We notice the
symmetry $g(\varepsilon,V)=g(-\varepsilon,V)$ and the fact that, through
(\ref{k-nu-eps2}), the transmission $T_{\nu}(\varepsilon)$ does depend on $V$.
In practice, we are only interested in the case $\varepsilon=eV$. Thus,
hereafter we refer to both $G$ and $g$ as functions of a single argument
$\varepsilon$ which is to be identified with $eV$ in the sense indicated in
Eqs. (\ref{g-gene}) and (\ref{reduced-g}).

The structure of the angular distribution of the conductance reflects that of
$g_{\nu}$ as a function of $\nu$, which generally reveals a complex and rich
behavior, since it is determined by the combined role of the product $T_{\nu
}(\varepsilon)T_{\nu}(-\varepsilon)$ and the cosine term in (\ref{reduced-g}).
Below we discuss some general trends.

First we note that $g_{\nu}(0)=T_{\nu}^{2}(0)/[2-T_{\nu}(0)]^{2}$, with
$T_{\nu}(0)$ computed for $eV=0$, which is consistent with Eq. (\ref{been}).
If the one-electron (normal) resonance occurs at a perpendicular energy
$E_{x}=E_{R}$ satisfying $\mu_{N}-E_{F}<E_{R}<\mu_{S}$, for $|\varepsilon
|<\Delta$ there is always a transverse mode $\nu(\varepsilon)$ for which
\begin{equation}
\mu_{S}-\epsilon_{\nu(\varepsilon)}+\varepsilon=E_{R}\ , \label{mu-nu-R}%
\end{equation}
i.e. such that $T_{\nu}(\varepsilon)$ presents a peak at $\nu=\nu
(\varepsilon)$ as a function of $\nu$, with maximum value $T_{0}$ (normal
resonance). In a symmetric structure, $T_{0}=1$.

As a function of $\nu$, the phases $\varphi_{\nu}(\pm\varepsilon)$ undergo an
abrupt change near $\nu(\pm\varepsilon)$, so that the cosine term goes quickly
through two maxima, in $\nu(\varepsilon)$ and $\nu(-\varepsilon)$, none of
which necessarily reaches unity. These maxima coincide in general with the
peaks of $T_{\nu}(\varepsilon)$ and $T_{\nu}(-\varepsilon)$. From
(\ref{reduced-g}), this translates into pairs of close lying peaks in the
conductance angular distribution. We have observed that the above tendency is
typically present for all intermediate values of $\varepsilon$ (as compared
with $\Delta$) for $(Z_{1},Z_{2})=(4,2)$ and $(0,2)$. Now we describe another
aspect of the peak formation mechanism that is relevant for $\varepsilon$ not
much smaller than $\Delta$ in the structure $(0,2)$. We note that it is
compatible with the trend discussed above.

Andreev resonances are characteristically given by the condition
\cite{leso97}
\begin{equation}
\cos\left[  \varphi_{\nu}(\varepsilon)-\varphi_{\nu}(-\varepsilon
)-2\vartheta(\varepsilon)\right]  =1~. \label{resonance}%
\end{equation}
If we recall that $\varepsilon$ is to be identified eventually with $eV$, and
that through (\ref{k-nu-eps2}) $\varphi_{\nu}(\varepsilon)$ does also depend
on $V$, we may state that, for a continuous range of voltages $V$, there is
always at least a value of $\nu=\bar{\nu}(V)$ satisfying (\ref{resonance}). As
defined in (\ref{mu-nu-R}), $\nu(0)$ is also a function of voltage, since
$\mu_{S}=\mu_{N}+eV$ with $\mu_{N}$ fixed. Alternatively, one may take
$\mu_{S}$ as fixed and $\mu_{N}$ dependent on voltage; then, $\nu(0)$ is
independent of $V.$ In both scenarios (and, conceivably, in intermediate
ones), there is a discrete set of values $\{V_{n}\}$ for which the two
transverses modes coincide, i.e. , for which $\bar{\nu}(V_{n})=\nu(0)$.

We note on the other hand that, for $\varepsilon=0,$ (\ref{mu-nu-R}) may also
be regarded as the maximum condition for $T_{\nu(0)}(\varepsilon)$ viewed as a
function of $\varepsilon$ with its maximum lying at $\varepsilon=0$ (i.e. at
total energy $\mu_{S}$). Thus we can assert that $T_{\nu(0)}(\varepsilon
)=T_{\nu(0)}(-\varepsilon)$ within a range of $\varepsilon$ values, which may
include $\varepsilon_{n}\equiv eV_{n}$. Noting that the Andreev resonance
condition (\ref{resonance}) is symmetric in $\varepsilon$, we conclude from
(\ref{reduced-g}) that
\begin{equation}
g_{\nu(0)}(\varepsilon_{n})=1~, \label{Andreev}%
\end{equation}
\textit{even }if $T_{\nu(0)}(\varepsilon_{n})$ is not unity. This maximum
value of the conductance per mode (which is 2 in units of $2e^{2}/h$; see Ref.
\onlinecite{sols99} for a discussion) is consistent with the results reported
in the single-mode study of Ref. \onlinecite{ried93}. Therefore, at voltages
$\{V_{n}\}$ the total transmission (summed over $\nu$) receives a strong
contribution from $\nu=\nu(0)$ and its vicinity. This behavior tends to
generate peaks in the total spectral conductance $G(\varepsilon)$ at or near
the values $\varepsilon_{n}=eV_{n}$ defined above.

The conclusion is that the sharpest resonances nucleate at angles near normal
resonances ($\nu(\varepsilon)$ is typically close to $\nu_{R}$, since
$|\varepsilon|<\Delta\ll E_{F}$). This happens for all energies $\varepsilon$.
However, as explained above, some energies $\varepsilon$ benefit more
efficiently from the resonance (in the sense that $g_{\nu}(\varepsilon)$
displays higher maximum values as a function of $\nu$) and thus give rise to
peaks in $G(\varepsilon)$ when integrated over angles.

Now we may argue like in Section III. Whenever $\mu_{S}>E_{R}$, there is a
low-lying transverse mode $\nu$ satisfying (\ref{mu-nu-R}). Then we expect to
have a strong peak in the \textit{angular distribution of the spectral
conductance}, $G(\varepsilon,\theta)$, which is defined to yield%

\begin{figure}
[ptbptbptbptbptb]
\includegraphics[width=8cm]{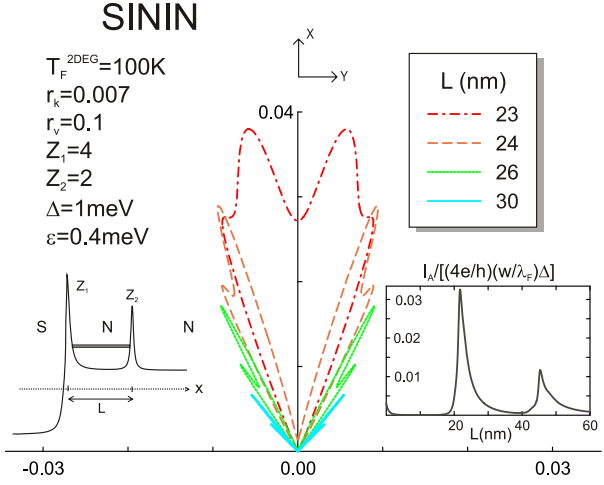}
\caption{(Color online) Normalized angular distribution of the
spectral conductance evaluated at $\varepsilon=0.4$ meV, for several
values of the interbarrier distance $L$.
The superconductor gap is $\Delta=$ 1 meV. Barrier strengths are $(Z_{1}%
,Z_{2})=(4,2)$. Left inset: schematic representation of the
perpendicular potential profile. Right inset: total current,
integrated over angles and energies (up to $\Delta$), normalized to
half its maximum possible value, as a
function of $L.$}%
\label{leso1}%
\end{figure}

\begin{figure}
[ptbptbptbptbptbptb]
\includegraphics[width=8cm]{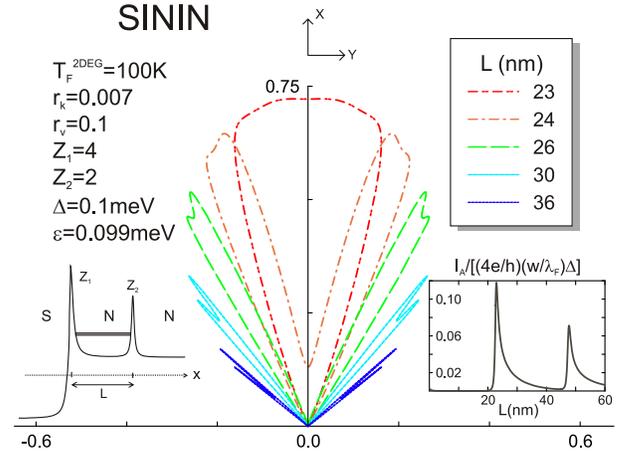}
\caption{(Color online) Same as Fig. \ref{leso1}, for $\Delta=0.1$
meV and $\varepsilon
=0.099$ meV.}%
\label{leso2}
\end{figure}

\begin{figure}
[ptbptbptbptbptbptbptb]
\includegraphics[width=8cm]{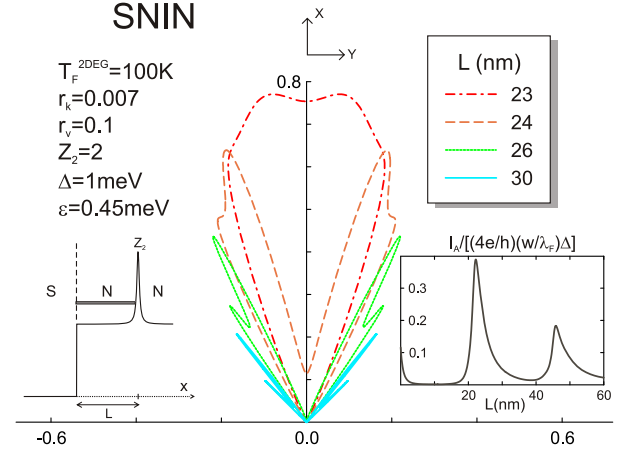}
\caption{(Color online) Same as Fig. \ref{leso1}, for
$(Z_{1},Z_{2})=(0,2)$ and
$\varepsilon=0.45$ meV.}%
\label{leso3}
\end{figure}

\begin{equation}
G(\varepsilon)=\int_{-\pi/2}^{\pi/2}d\theta~G(\varepsilon,\theta)\ .
\label{ang-distr-eps}%
\end{equation}
Figs. 5-7 show the normalized value of $G(\varepsilon,\theta)$ for structures
with $(Z_{1},Z_{2})=(4,2)$ and $(0,2)$, the former being considered for two
different combination of $\varepsilon$ and $\Delta$. As $L$ increases, the
value of $E_{R}$ decreases and sinks below $\mu_{S}$. This generates maxima in
the angular distribution in the manner discussed above.

At zero temperature, and for $eV>0$,$\ G(\varepsilon)$ can be understood as
the contribution to the total current stemming from electron pairs emitted
into the normal metal with total energies $\mu_{S}\pm\varepsilon$. The two
electrons leaving the superconductor have identical $\left\vert k_{x}%
\right\vert $ and slightly different total energy (see below). Thus they do
not point exactly in the same direction, i.e. the V which they form upon
emission is not exactly centered around the normal axis. By symmetry, for each
pair in which e.g. the upper electron is emitted towards the right (and the
lower one to the left), there is another pair solution in which the upper
electron travels to the left (and the lower one to the right). When plotting
the total differential conductance, the two asymmetric Vs appear as a single V
whose lobes are double peaked.

We note here that, in the contribution to \thinspace$G(\varepsilon)$ as
defined in Eqs. (\ref{g-gene}) and (\ref{reduced-g}), $T_{\nu}(-\varepsilon)$
is identical to the $T_{\nu}$ appearing in the zero voltage limit discussed in
the previous section [see Eq. (\ref{k-nu-eps2})], i.e. $\nu(-\varepsilon
)=\nu_{R}$ as defined in (\ref{resonancia-normal}), if we identify $\mu
_{N}\equiv\mu$. This implies that, in the double-peaked lobes, the inner peak
points in the same direction as the single-peaked lobe of the linear ($V=0$)
limit, a result which is independent of the sign of $eV$. The fact that the
coincidence occurs at the inner peak can be understood by noting that, since
$\varepsilon=eV$, we have $k_{\nu(\varepsilon)}(\varepsilon)=k_{\nu
(-\varepsilon)}(-\varepsilon)$, while $\epsilon_{\nu(\varepsilon)}%
=\epsilon_{\nu(-\varepsilon)}+2eV$. Thus, at a given $\varepsilon$, peaks in
the angular distribution occur at $\nu(\varepsilon)$ and $\nu(-\varepsilon)$.
Both have the same perpendicular momentum, but the latter has lower parallel
kinetic energy.

The fact observed in Figs. 5-7 that the inner peak displays a larger current
density is due to the asymmetric character of the peaks in $T_{\nu}%
(\pm\varepsilon)$ as a function of $\nu$ (or the angle $\theta$),
which ultimately reflects the greater efficiency with which
close-to-normal emission electrons contribute to the electric
current.

The insets of Figs. 5-7 show the total current (integrated over $\theta$ and
$\varepsilon$) as a function of $L$. As for the zero bias conductance, they
reveal a succession of maxima followed by an inverse square root decay law
that mirrors the transverse density of states (see discussion in the previous section).

\begin{figure}
[ptbptbptbptbptbptbptbptb]
\includegraphics[width=8cm]{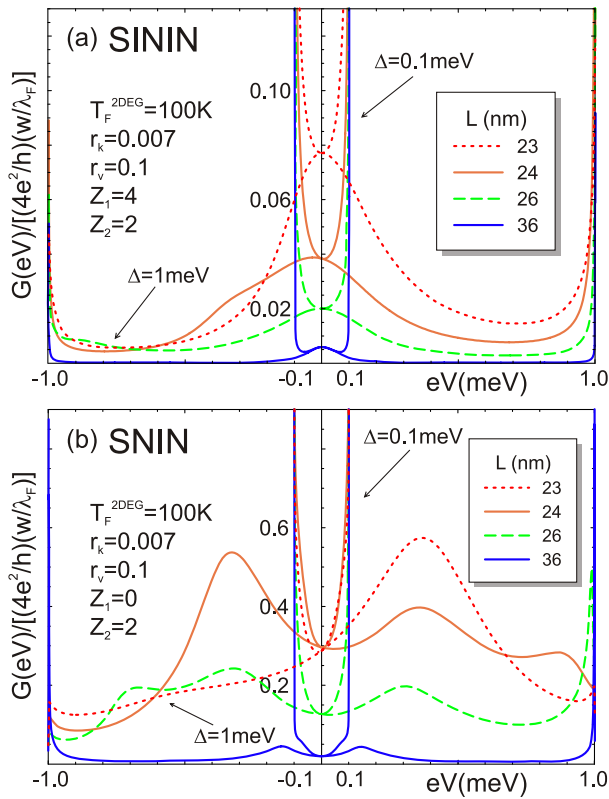}
\caption{(Color online) Subgap spectral conductance (integrated over
angles), normalized to half its maximum possible value, for four
values of the interbarrier distance $L$, two values of the
superconductor gap $\Delta$, and two values of the
strength $Z_{1}$ of the barrier located right at the SN interface.}%
\label{clasi}
\end{figure}

Fig. 8 shows the total spectral conductance for voltages below the
gap. This type of curves has been the object of preferential
attention in the previous literature on NS transport. By presenting
them here, we make connection with that preexisting body of
knowledge, in particular with the experimental and theoretical works
of Refs. \onlinecite{giaz03} and \onlinecite{leso97}, respectively.
The forthcoming remarks are intended to complement that discussion
and to provide a self-contained, unified picture of the work
presented here.

The asymmetry in $G(eV)$ is due to the finite normal bandwidth. For the
results plotted in Fig. 8, the voltage $V$ varies as $\mu_{S}$ varies with
$\mu_{N}$ fixed. From Fig. 1, it is clear that raising $\mu_{S}$ is not
equivalent to lowering it. Asymmetric curves are displayed in Ref.
\onlinecite{choi05} and have been discussed in Ref. \onlinecite{leso97} (see
also references therein). In what follows, we focus on the behavior for $eV>0$.

Both in Figs. 8a and 8b we present two groups of curves, corresponding to a
small and a large gap. The barrier parameters of Fig. 8a are the same as those
of Figs. 5 and 6 , namely, $(Z_{1},Z_{2})=(4,2)$. Although Figs. 5 and 6
already exhibit Andreev features such as the double-peaked lobes in
$G(\varepsilon,\theta)$, these are washed out when the angular variable is
integrated to yield the total spectral conductance $G(\varepsilon=eV)$, as
shown by the single peaked curves obtained for the same value of the gap as in
Fig. 5 ($\Delta=1$ meV), or by the absence of peaks for the parameters of Fig.
6 ($\Delta=0.1$ meV$)$. The curves for $\Delta=1$ meV display a clear zero
bias conductance peak (ZBCP) whose height is determined by the structure
normal properties [see Eq. (\ref{been})]. As $\varepsilon$ increases above
zero, both electron and holes (or both the upper and lower energy emitted
electrons) may benefit from the low-lying normal resonance ($E_{R}<\mu_{S}$)
as long as $\varepsilon<\Gamma$, where $\Gamma$ is the linewidth of the normal
resonance. When $\varepsilon>\Gamma$, it is not possible to channel both
electrons through the same resonance and the contribution to the conductances
decreases. On closer inspection, one finds that the width of the ZBCP is
indeed determined by the normal resonance width, but not by that appearing in
the perpendicular transmission $T_{\nu}(\varepsilon)$ (viewed as a function of
$\varepsilon$). Rather, it essentially mirrors the width of the numerator in
Eq. (\ref{reduced-g}). This is the product $T_{\nu}(\varepsilon)T_{\nu
}(-\varepsilon)$ evaluated at $\nu_{R}$ and viewed also as a function of
$\varepsilon$, i.e. for electrons leaving in the direction of maximum current
flow (at exit angle $\theta=\theta_{R}$). This is reminiscent of the result
stating that, when $Z_{2}$ is replaced by a disordered normal metal, the width
of the ZBCP is of the order of the Thouless energy \cite{leso97}.

A general property of SN interfaces with a single barrier right at
the interface is that Andreev reflection probability tends to unity
as $\left\vert \varepsilon\right\vert \rightarrow\Delta$
\cite{blon82}. However, we find that this is generally not the case
for a double barrier interface. For $Z_{1}=0$, we do notice that
sharp peaks in $G(\varepsilon)$ form just below the gap for some
values of $L$, so close to it that they can be observed only through
a magnification of Fig. 8. Due to this tendency to acquire large
values near the gap, $G(\varepsilon)$ goes through a minimum at
finite $\varepsilon$ if the width of the ZBCP is smaller than the
gap. This is the case shown in Fig. 8a for $\Delta=1$ meV. For a
smaller gap ($\Delta=0.1$ meV), the value of $G(0)$ remains
unchanged but there is no room for $G(\varepsilon)$ to display a
minimum between 0 and $\Delta$.

Being $Z_{1}=0$ more transmissive (although not entirely, because of the
reflection at the potential step; see Section III), Fig. 8b displays Andreev
resonance features that do survive upon integration over angles. For
$\Delta=1$ meV and $L=$ 23 nm, one observes a peak at finite energies that
adds to the overall ZBCP. As $L$ increases, the inner Andreev peak evolves
towards zero energy. At larger distances ($L=$ 36 nm), the lowest Andreev
resonance can only be hinted at as a shoulder in the plot for $\Delta=0.1$
meV. We also note that, for $L=$ 24 and 26 nm, a second Andreev resonance
becomes visible close to the gap edge. \ However, due to the involved
interplay between the transmission probabilities and the cosine term appearing
in Eq. (\ref{reduced-g}), this second peak does not appear to follow a simple
monotonic trend. In fact, for $\Delta=1$ meV, the second resonance is no
longer observable because it evolves towards a sharp peak just below the gap.

\section{Discussion}

\label{discussion}

So far we have assumed that the SN interface is infinitely long
($w\rightarrow\infty$). This has allowed us to treat $k_{y}$ as a
continuous, conserved quantum number, which considerably simplifies
the transport calculation. Of course, the idea of an infinite
interface is at odds with the primary motivation of our work, which
is to propose a method to spatially separate mutually entangled
electron beams. Below we argue that, fortunately, only a moderately
long interface is needed in practice.

For simplicity, we focus our discussion on the low voltage limit, where the
total energy can be assumed to be sharply defined. Then the width
$\Delta\theta$ of the angular distribution is due only to the uncertainty in
the parallel momentum $\Delta k_{y}$. This in turn is closely connected to
$\Delta k_{x}$ through the relation $k_{x}\Delta k_{x}=k_{y}\Delta k_{y}$,
since total energy uncertainty is zero. There are two contributions to the
momentum uncertainty: the nonzero width of the resonance in the perpendicular
transmission and the finite length of the SN interface. Thus we may estimate%
\begin{equation}
\Delta k_{y}\simeq\frac{m\Gamma}{\hbar^{2}k_{y}}+\frac{1}{w}\ .
\label{Delta-ky}%
\end{equation}
This translates into an angular width%
\begin{equation}
\Delta\theta\simeq\frac{\Gamma}{E_{F}\sin(2\theta_{R})}+\frac{1}{k_{F}%
w\cos\theta_{R}}\ . \label{delta-th}%
\end{equation}
The actual angular width of $G(0,\theta)$ is actually a bit smaller, since the
present estimate is based on one-electron considerations, while the relevant
angular distribution is determined by Eq. (\ref{been}). We neglect this
difference for the present simple estimates.

Eq. (\ref{delta-th}) contains two contributions. The first term is
determined by the normal resonance and is responsible for the width
of the angular distributions plotted in Figs. 2-4 (with
$w\rightarrow\infty$). Our main concern here is that the second
contribution, that which stems from the finiteness of the aperture,
does not contribute significantly.

A strict criterion may be that the interface finite length should
not modify the intrinsic angular width ($\hbar
v_{F}\sin\theta_{R}/w\ll\Gamma$), which everywhere has been assumed
to be small enough to allow for narrow divergent beams. A more
lenient criterion is that, regardless of the specific value of
$\Gamma$, the finite aperture should not generate an excessively
broad angular distribution. For typical cases this amounts to
requiring $k_{F}w\gg1$ (for a discussion see Fig. 5 in Ref.
\onlinecite{prad04}). For the bandwidth which we
have assumed ($E_{F}/k_{B}=100$ K) and an effective mass of $m=(r_{k}%
/r_{v})m^{\prime}=0.07m_{e}$, where $m_{e}$ is the bare electron mass, we have
$\lambda_{F}=2\pi/k_{F}\sim50$ nm. So apertures greater than a few hundred
nanometers seem desirable to keep the angular uncertainty within acceptable bounds.

Another source of angular spreading is interface roughness, with a
characteristic length scale $l$. However, it should not pose a fundamental
problem as long as $l\gg\lambda_{F}$, so that a structure of intermediate
width could be designed satisfying $l\gg w\gg\lambda_{F}$.

For the difference in velocity direction to translate into spatial separation,
it is necessary that the spin detectors are placed sufficiently away from the
electron-emitting SN interface. Of course, the needed distance depends also on
the exit angle $\theta_{R}$. For a convenient value of $\theta_{R}\sim\pi/4$,
simple geometrical considerations suggest that, unsurprisingly, the distance
$d$ from the detector to the center of the SN interface must be greater than
its width $w$. Since elastic mean free paths in a 2DEG can be made as high as
$l_{e}\sim$100 $\mu$m, there seems to be potentially ample room for building
structures satisfying $\lambda_{F}\ll w\ll d\ll l_{e}$. Such devices would
display well-defined divergent current lobes which could be detected (and,
eventually, manipulated) at separate locations before the directional focusing
is significantly reduced by elastic scattering.

\section{Conclusions}

\label{conclusions}

We have investigated theoretically the possibility of creating hybrid
normal-superconductor structures where the two electrons previously forming a
Cooper pair in the superconductor are sent into different directions within
the normal metal. The central idea relies on the design of a structure that is
transparent only to electrons with perpendicular energy within a narrow range
of a resonant level. Since the total energy lies close to the Fermi level,
such a filtering of the electron perpendicular energy translates into
\textit{exit angle selection}.

Electrons from a conventional superconductor are known to be correlated in
such a way that electrons moving at similar speeds in opposite directions tend
to have opposite spin. At low temperatures and voltages, electron flow from
the superconductor to the normal metal is entirely due the transmission of
correlated electron pairs. These have both opposite spin and opposite parallel
(to the interface) momentum, while possessing the same total energy. If the
exit angle is selected by filtering the perpendicular momentum, the current in
the normal metal is formed by two narrow, mutually singlet entangled electron
beams which point in different directions and which spatially separate from
each other at distances from the source much greater than the width of the source.

The trick of exit angle selection is intended to facilitate a neat observation
of nonlocal entanglement between electron beams, and this article has been
devoted to proposing a specific implementation of that idea. One cannot help
noting, however, that such a selection of the outgoing direction might not be
totally essential. If we content ourselves with measuring anticorrelated
low-energy spin fluctuations over mesoscopic length scales, it may just be
sufficient to place the two spin detectors symmetrically around the interface
at a sufficient distance and angle, very much like in the setup of Fig. 1a but
with a conventional, non angle-selecting SN tunnel interface. If their motion
between the emitter and the detector is ballistic, electrons arriving at each
detector have, on average, opposite parallel momentum and opposite spin
(angular anticorrelation has been explicitly shown in Ref. \onlinecite{prad04}
for a broad perfect interface). The boundaries of the 2DEG might conceivably
be designed to optimize such correlations. The outcome is that electrons
arriving at each detector will exhibit a degree of nonlocal spin-singlet
correlations that could be measured.

Altogether, we conclude that a ballistic two-dimensional electron
gas provides an ideal scenario to probe nonlocal entanglement
between electrons emitted from a distant, finite-size interface with
a superconductor. If that interface is formed by a hybrid structure
that selects the perpendicular energy and thus the magnitude of the
electron exit angle, nonlocal spin correlations will be clearly
observed if the outgoing beams are directed towards suitably placed
detectors.\bigskip

\section*{Acknowledgments}

This research has been supported by MEC (Spain) under Grants No. BFM2001-0172
and FIS2004-05120, the FPI Program of the Comunidad de Madrid, the EU Marie
Curie RTN Programme under Contract No. MRTN-CT-2003-504574, and the Ram\'{o}n
Areces Foundation.

\end{document}